# ON SOLVING THE CORONAL HEATING PROBLEM

## JAMES A. KLIMCHUK[1]

[1]*Space Science Division, Naval Research Laboratory, Washington, DC 20375, USA*

*klimchuk@nrl.navy.mil*

# ON SOLVING THE CORONAL HEATING PROBLEM

## JAMES A. KLIMCHUK[1]

*[1]Space Science Division, Naval Research Laboratory, Washington, DC 20375, USA*

(29 November 2005)

**Abstract.** The question of what heats the solar corona remains one of the most important problems in astrophysics. Finding a definitive solution involves a number of challenging steps, beginning with an identification of the energy source and ending with a prediction of observable quantities that can be compared directly with actual observations. Critical intermediate steps include realistic modeling of both the energy release process (the conversion of magnetic stress energy or wave energy into heat) and the response of the plasma to the heating. A variety of difficult issues must be addressed: highly disparate spatial scales, physical connections between the corona and lower atmosphere, complex microphysics, and variability and dynamics. Nearly all of the coronal heating mechanisms that have been proposed produce heating that is impulsive from the perspective of elemental magnetic flux strands. It is this perspective that must be adopted to understand how the plasma responds and radiates. In our opinion, the most promising explanation offered so far is Parker's idea of nanoflares occurring in magnetic fields that become tangled by turbulent convection. Exciting new developments include the identification of the "secondary instability" as the likely mechanism of energy release and the demonstration that impulsive heating in sub-resolution strands can explain certain observed properties of coronal loops that are otherwise very difficult to understand. Whatever the detailed mechanism of energy release, it is clear that some form of magnetic reconnection must be occurring at significant altitudes in the corona (above the magnetic carpet), so that the tangling does not increase indefinitely. This article outlines the key elements of a comprehensive strategy for solving the coronal heating problem and warns of obstacles that must be overcome along the way.

## 1. Introduction

Stated simply, the coronal heating problem is to identify and understand the physical mechanism responsible for heating the corona to multi-million degree temperatures, several hundred times hotter than the underlying photosphere. The problem first became apparent more than six decades ago, when Grotrian (1939) and Edlen (1942) realized that emission lines seen during total solar eclipses were not due to a new element dubbed "coronium," but rather to known elements at very high stages of ionization. The extreme temperatures necessary for such ionization have puzzled solar physicists ever since. There are no doubt many different heating mechanisms operating in the corona, and the real goal is to determine which one is dominant, both in general and in specific situations. A number of important questions must be answered. Are distinct coronal loops heated differently from the diffuse corona? Are there different classes of loops that are heated in different ways? Is quiet Sun heating similar to active region heating? Are stellar coronae heated in the same way as the solar corona?

A number of plausible theories for coronal heating have been proposed, but it has proved difficult to determine which ones, if any, are actually correct. Some of the theories are really not much more than conceptual ideas. Others are far better developed, but nonetheless rely on modeling that is highly idealized or incomplete. This inadequate state of affairs should not be taken as an indictment of the discipline. Rather, it reflects the extreme difficulty of the problem. A definitive test of any coronal heating theory requires a quantitative prediction of observable quantities that is based on a detailed, first principles treatment. This is a monumental task, as we will show.

It is instructive to think of the coronal heating problem as involving a number of separate steps, as indicated in the "coronal heating flowchart" of Figure 1. We must identify a source of energy and a mechanism for converting that energy into heat. We must determine how the plasma responds to the heating. Finally, we must predict the spectrum of emitted radiation and its manifestation as observable quantities. Only by accomplishing *all* of these steps can a truly rigorous and meaningful comparison with actual observations be made.

Most coronal heating studies have focused on restricted parts of the flowchart, largely because of the unique challenges involved in each step. Multi-dimensional MHD models are typically used to study the source and conversion of energy, but they provide little useful information on the response of the plasma and its radiation signatures. Conversely, 1D hydrodynamic models are typically used to study the plasma response, but they treat the heating in a generic manner, without regard to its physical origin. Both approaches are valuable, but it is dangerous to apply them in isolation without considering the big picture. An important goal of this article is to encourage a more integrated approach to the coronal heating problem and to emphasize the importance of all of the steps in the flowchart. Any missing or weak links can lead to a false sense of success.

We address each of the flowchart steps in the next five sections. Particular emphasis is given to energy conversion and plasma response, since these are in many ways the most challenging. Several cross-cutting themes emerge: (1) highly disparate spatial scales, (2) physical connections between the corona and lower atmosphere, (3) complex microphysics, and (4) variability and dynamics. We then offer our personal view of the most promising explanation for how the magnetically-closed corona is heated. It is essentially Parker's (1983, 1988) idea of nanoflares occurring in tangled magnetic fields, as elaborated upon recently by Priest, Heyvaerts, and Title (2002). We show that a mechanism called the secondary instability is likely to be responsible for the impulsive energy release, and that impulsive heating is able to explain certain observed properties of coronal loops that are otherwise difficult to understand. We close with a summary and recommendations.

This article is not intended to be an exhaustive review of coronal heating. For such information, the reader is referred to Gómez (1990), Zirker (1993), Narain and Ulmschneider (1996), Mandrini, Démoulin, and Klimchuk (2000), Walsh and Ireland (2003), and Aschwanden (2004). We deal explicitly with the magnetically-closed corona, where field lines are rooted to the solar surface at both ends. Some of the discussion may also apply to the magnetically-open solar wind, but there are many other important aspects of solar wind heating that we do not address. We also do not address the heating of the chromosphere, which is an equally important problem. This article stems from a keynote address given at the highly successful SOHO-15 Workshop on Coronal Heating that was held in St. Andrews, Scotland in the fall of 2004. An early version of the article appears in the proceedings (Klimchuk, 2004).

## 2.  Energy Source

The combined radiative and conductive energy losses from the corona have a flux of roughly $10^7$ ergs cm$^{-2}$ s$^{-1}$ in active regions and 3x10$^5$ erg cm$^{-2}$ s$^{-1}$ in the quiet Sun (Withbroe and Noyes, 1977). A basic requirement of any coronal heating theory is to identify an energy source that can sustain these levels of losses.  It is widely accepted that mechanical motions in and below the photosphere are the ultimate source of the energy.  These motions displace the footpoints of coronal magnetic field lines and either quasi-statically stress the field or generate waves depending on whether the timescale of the motion is long or short compared to the end-to-end Alfvén travel time.  Dissipation of magnetic stresses is referred to as direct current (DC) heating, and dissipation of waves is referred to as alternating current (AC) heating.  We examine each of these in turn.

### 2.1.  DC Heating

Footpoint motions perform work on the coronal magnetic field and increase its free energy at a rate given by the Poynting flux through the base:

$$\boldsymbol{F} = -\,\frac{1}{4\pi}\,B_v\,\boldsymbol{B_h}\cdot\boldsymbol{V_h}\,,\tag{1}$$

where $B_v$ and $\boldsymbol{B_h}$ are the vertical and horizontal components of the field and $\boldsymbol{V_h}$ is the (horizontal) footpoint velocity.  The energy of the field will also change if there is emergence or submergence of flux, but we do not consider those processes here.

Most of the magnetic flux in the photosphere is concentrated in small tubes of kilogauss strength (e.g., Solanki, 1993; Muller, 1994; Socas-Navarro and Sánchez Almeida, 2002; Domínguez Cerdeña, Kneer, and Sánchez Almeida, 2003).  These flux tubes flare out in the chromosphere and transition region and become space filling in the corona where the plasma $\beta$ (= $8\pi P/B^2$) is small.  This occurs at very low altitudes within active regions and the quiet Sun network. Within mixed polarity inter-network regions, a web of small loops forms a low-lying "magnetic carpet" (Schrijver and Title, 2002).  Most of the carpet loops do not penetrate into the corona (Close et al., 2003), and β may be of order unity within some of those that do (Schrijver and van Ballegooijen, 2005).  A fraction of the inter-network flux extends above the carpet and spreads out in the higher corona where β is small.  The quiet Sun coronal field is therefore a mixture of network field and surviving inter-network field.

To evaluate Equation (1), it is convenient to take the base of the corona to be the lowest altitude where the expanding flux tubes have merged.  Photospheric longitudinal magnetograms with modest spatial resolution provide a good estimate of the vertical field strength at this altitude. The signal measured near disk center is indicative of the net flux in an observational pixel and, for coarse pixels, is insensitive to the small loops of the magnetic carpet.  Observations give $B_v$ flux densities of typically 100 G in active region plage areas (Schrijver and Harvey, 1994) and 5-10 G in the quiet Sun (4" pixels; López Fuentes, private communication).

Magnetic flux tubes in the photosphere are displaced by turbulent convection and are observed to wander about the surface with a characteristic velocity $V_h$ of order 1.0x10$^5$ cm s$^{-1}$ (Muller *et al.*, 1994; Berger and Title, 1996).   We can assume that the tubes move with similar velocities at the coronal base just above.  Using these values in Equation (1) and assuming that $B_h \sim B_v$, we find that the Poynting flux into the corona is adequate to explain the observed energy losses of both the quiet Sun and active regions.  There is no question that DC heating is viable from an energy source standpoint. Understanding how the magnetic stress energy is converted into heat is a much bigger challenge. We will return to Equation (1) in Section 3.3 to infer a fundamental property of the energy conversion process.

## 2.2. AC HEATING

The energetic feasibility of AC heating is much less certain. The same turbulent convection that quasi-statically stresses the coronal field also generates a large flux of upwardly propagating waves. These waves take on a variety of different forms: acoustic, Alfvén, and fast and slow magnetosonic plane waves, as well as torsional, kink, and sausage magnetic flux tube waves (body and surface varieties). Mode coupling and other processes transfer energy between the different wave types, so the mix of waves changes as a function of height in the atmosphere (e.g., Stein and Nordlund, 1991; Bogdan et al., 2003). Theoretical and observational estimates suggest energy fluxes at the top of the convection zone of roughly several times $10^7$ erg cm$^{-2}$ s$^{-1}$ (Narain and Ulmschneider, 1996).

Energy fluxes of this magnitude are more than adequate to heat the corona. However, only a small fraction of the flux is able to pass through the very steep density and temperature gradients that exist in the chromosphere and transition region. Acoustic and slow-mode waves form shocks and are strongly damped, while fast-mode waves are strongly refracted and reflected (Narain and Ulmschneider, 1996). Alfvén waves, including the Alfvén-like torsional and kink tube waves, are best able to penetrate into the corona. They do not form shocks since they are transverse, and their energy is ducted along the magnetic field rather than being refracted across it. Energy flux estimates for these waves are therefore especially important. Ulrich (1996) observed magnetic and velocity fluctuations with the correct phase relationship for Alfvén waves and inferred a flux of $\leq 10^7$ erg cm$^{-2}$ s$^{-1}$ in regions of strong magnetic field. This is marginally adequate to heat active regions, but only if the transmission efficiency is near 100%. Muller et al. (1994) inferred a similar energy flux for the quiet Sun from the observed "shaking" of photospheric flux tubes by granular motions. Only a small fraction of this flux needs to reach the corona to heat the quiet Sun. On a negative note, Parker (1991) has argued that solar convection should not be very efficient at generating Alfvén waves (see also Collins, 1992).

Most Alfvén waves are strongly reflected in the chromosphere and transition region, where the Alfvén speed increases dramatically with height. Significant transmission is possible only within narrow frequency bands centered on discrete values where loop resonance conditions are satisfied (Hollweg, 1981, 1984; Ionson, 1982). Hollweg (1985) estimates that enough flux may pass through the base of long (> $10^{10}$ cm) active region loops to provide their heating, but shorter loops are a problem, since they have higher resonance frequencies and the photospheric power spectrum is believed to decrease exponentially with frequency in this range. Litwin and Rosner (1998) suggest that short loops may in fact transmit waves with low frequencies, as long as the field is sufficiently twisted.

Transmission is not an issue for waves that are generated in the corona itself. Possible sources include magnetic reconnection (Moore et al., 1991; Yokoyama, 1998; Sturrock, 1999; Longcope, 2004) and loss of equilibrium (Longcope and Sudan, 1992). The energy for these waves first resides in the stressed coronal magnetic fields, so heating scenarios of this type have both AC and DC aspects. The ability of waves to heat regions that are remote from DC energy conversion sites could be very important. For example, one might imagine that magnetic reconnection produces localized strong heating in a bright coronal loop and more distributed weak heating in the surrounding diffuse corona.

The existence of waves in the corona is well established (e.g., Erdélyi, Ballester, and Fleck, 2004). For instance, traveling intensity perturbations observed in the legs of some loops provide clear evidence for upwardly propagating acoustic waves (e.g., Berghmans and Clette, 1999; De Moortel *et al.*, 2002). The real issue is whether coronal waves carry a sufficient energy flux to heat the plasma. The spatially-resolved acoustic waves most certainly do not. Much larger energy fluxes may be present in unresolved wave motions. The observed non-thermal broadening of transition region and coronal spectral lines implies fluxes that are sufficient to heat both the quiet Sun and active regions, but it is completely unknown whether the waves are of the right type or the right frequency to be adequately damped (Porter, Klimchuk, and Sturrock, 1994). Furthermore, the non-thermal line broadening could be produced by unresolved motions that are completely unrelated to waves, such as field-aligned flows that are driven by impulsive heating (e.g., Patsourakos and Klimchuk, 2005b). Our lack of knowledge about the basic properties of coronal waves (type, energy flux, and spectrum) is a major impediment to understanding AC coronal heating.

## 3. Energy Conversion

### 3.1. Disparate Spatial Scales

The next step in the coronal heating flowchart is to identify a mechanism for converting the magnetic stress energy or wave energy into heat. Because classical dissipation coefficients are extremely small in the corona, significant heating generally requires the formation of very steep gradients and very small spatial scales. Magnetic gradients and their associated electrical currents lead to heating by reconnection and Ohmic dissipation, while velocity gradients lead to heating by viscous dissipation.

There are a number of different ways to produce the necessary gradients. They can be formed through slow quasi-static evolution and through highly dynamical processes. Possible scenarios include: (1) simple photospheric flow patterns at the base of complex coronal magnetic fields containing separatrix surfaces, nulls, and separators (e.g., Karpen, Antiochos, and DeVore, 1996; Longcope, 1996, 1998; Aly and Amari, 1997; Priest, Heyvaerts, and Title, 2002); (2) complex photospheric flow patterns with stagnation points at the base of simple coronal magnetic fields (e.g., van Ballegooijen, 1986; Galsgaard and Nordlund, 1996; Hendrix *et al.*, 1996; Antiochos and Dahlburg, 1997); (3) instabilities, such as the kink instability brought about by a twisting flow (e.g., Hood and Priest, 1979; Lionello *et al.*, 1998); (4) turbulence (e.g., Inverarity, Priest, and Heyvaerts, 1995; Einaudi *et al.*, 1996; Dmitruk, Gómez, and DeLuca, 1998); and (5) loss of equilibrium, usually considered for flares and coronal mass ejections (e.g., Forbes and Isenberg, 1991), but also relevant to coronal heating (Longcope and Sudan, 1992). For waves, we have the additional processes of (6) resonance absorption (e.g., Ionson, 1978; Hollweg, 1984; Davila, 1987; Poedts, Goossens, and Kerner, 1989; Steinolfson and Davila, 1993), which we examine in more detail in Section 4.3; and (7) phase mixing (e.g., Heyvaerts and Priest, 1983; Similon and Sudan, 1989; Ofman and Aschwanden, 2002).

An important point is that all of these mechanisms for producing small-scale structure involve the large-scale magnetic field and its connection to the photosphere. The small scales develop either gradually or suddenly as the large-scale coronal field responds to slow motions at the lower boundary. Precisely how this occurs depends on the details of the field and the details of the motions (although turbulence is rather generic). This poses a severe challenge for modeling energy conversion, since an enormous range of scales must be treated. To date, no study has been fully successful. Some numerical simulations have accurately addressed the build up of stresses and the formation of current sheets, but the energy conversion is highly under-resolved and therefore questionable. Other simulations have attacked the details of the energy conversion with high resolution, but the existence and properties of a current sheet are simply assumed.

It is clear that a simultaneous treatment of all the important spatial scales requires a highly non-uniform numerical mesh. The mesh must also be adaptive, since the locations where steep gradients will form are not generally known in advance. Even when the initial locations can be anticipated, they usually drift as the simulation proceeds. Mesh points must therefore be dynamically added in places where they are needed and removed in places where they are no longer necessary. Techniques of this kind are now available and are beginning to produce impressive results for certain classes of problems (e.g., Gombosi *et al.*, 2004; Antiochos and DeVore, 2004). However, a number of challenges must still be overcome. More sophisticated algorithms for modifying the mesh must be developed for problems in which many small-scale structures are scattered throughout the computational domain (e.g., turbulence). Existing algorithms tend to produce a uniformly fine mesh in these situations, and the simulation grinds to a crawl. Another issue concerns the accuracy of the solution at the boundaries between sub-domains of differing mesh size. These problems are reasonably straightforward, and rapid progress can be expected in the near future.

## 3.2. MICROPHYSICS

Detailed microphysics is likely to play an important role in the energy conversion process. Anomalously large (non-classical) transport coefficients may be required for significant heating, even in the presence of steep gradients. It was argued by Parker (1973) and subsequently verified with numerical simulations by Biskamp (1993) and others that magnetic reconnection proceeds at the fast Petschek rate only if the electrical resistivity is enhanced by 3 or more orders of magnitude. In addition, the enhancement *must* be spatially localized and not spread over a large volume (see also Kulsrud, 2001).

There have been several attempts to infer the values of coronal transport coefficients by studying their effects on observed wave motions. Nakariakov *et al*. (1999) and Ofman and Aschwanden (2002) concluded that the rapid damping of oscillating *TRACE* loops requires at least a 5 order of magnitude enhancement in either the electrical resistivity or the shear viscosity (although Hollweg, 1984, argued that loop oscillations may dampen quickly from energy leakage out the ends). On the other hand, Klimchuk, Tanner, and DeMoortel (2004) found that classical thermal conduction is adequate to explain the damping of acoustic waves propagating along *TRACE* loops and ruled out the possibility that either thermal conduction or compressive viscosity are greatly enhanced. It may seem inconsistent that some transport coefficients would be anomalously enhanced while others not, but in fact this is not surprising. The magnetic field produces tremendous anisotropies in the particle transport, both for classical transport by collisions and anomalous transport by, for example, small-scale turbulence. Shear viscosity, which involves cross-field transport, is much more affected by turbulence than is field-aligned thermal conduction (Braginskii, 1965). Further quantitative study of these effects is needed.

Another microphysics issue concerns the collisionality of the plasma. Recent studies suggest that collisionless effects may be extremely important for reconnection (e.g., Birn *et al*. 2001; Bhattacharjee, 2004). The Hall term in the generalized Ohm's Law causes ions to decouple from the magnetic field whenever the current sheet is thinner than the ion skin depth, or $\sim 10^3$ cm in the corona (Priest and Forbes, 2000, Table 1.2). For reasons that are not yet entirely clear, this allows reconnection to proceed at the Petschek rate even if the resistivity is everywhere small. Conceptually, one can think of this in terms of a collisional plasma in which the resistivity is enhanced only at the sheet, thereby providing closure with the results of Parker and Biskamp cited above. It should be noted that collisionless reconnection studies have mostly been restricted to the neutral sheet configuration, which is a special type of current sheet that contains no guide field. It has not yet been established that collisionless reconnection is rapid in the more general case where the magnetic field rotates across the sheet, though very recent simulations suggest that it is (Drake *et al.*, 2005)

Hybrid codes that simulate both the MHD and particle aspects of the plasma are extremely powerful for investigating isolated coronal regions such as current sheets. However, they are limited to studies of very small regions, and we have emphasized that the interplay between large and small spatial scales is critically important for understanding coronal heating (and most other interesting coronal problems). Some modelers have argued that the details of the microphysics are unimportant, and that the artificial numerical resistivity of existing MHD codes captures the essential physical effects that determine how a coronal system will evolve. This may be a reasonable working hypothesis, given our present numerical limitations, but it is a dangerous position to adopt as an unchallenged truth. Let us examine this point more closely.

The numerical resistivity of a finite-difference simulation with mesh size $\Delta x$ and time step $\Delta t$ can be expressed as

$$\eta_{num} = f \frac{\Delta x^2}{\Delta t} \quad , \qquad (2)$$

where $f$ is a dimensionless function of both the local flow speed and the ratio $\Delta x/l$, $l$ being the local spatial scale of the magnetic field (C.R. DeVore, private communication). Note that $\eta_{num}$ has the same units as physical resistivity, length-squared over time. The magnitude and precise form of $f$ are algorithm dependent, but all modern, high-order algorithms have the generic property that $f$ is greatest where $\Delta x/l$ is largest. It seems intuitively correct that numerical effects should be greatest where the gradients are most poorly resolved. Furthermore, algorithms without this property tend to be unstable or produce solutions with unphysical oscillatory structure, so they are not generally used.

The point of this discussion is to show that numerical resistivity mimics the behavior of the corona in at least a gross sense. For a given mesh size, the resistivity is small away from current sheets, where $l$ is large, and it is large within current sheets, where $l$ is small. However, is the similarity with the actual corona close enough? Can we be confident that differences do not fundamentally alter the outcome of a simulation? In our opinion, we do not yet have a clear answer. At the very least, modelers should repeat their simulations with several different $\Delta x$ to verify that the results do not depend qualitatively on the numerical resolution (within the examined range). Questions remain even then, since the Reynolds numbers attainable with the best MHD simulations are still many orders of magnitude smaller than solar values, and there is no guarantee that the results can be scaled to realistic solar conditions. It is not far-fetched to imagine that the simulation is dominated by one process, but the real corona is dominated a completely different process that cannot be captured by the simulation. Our intent here is not to paint a pessimistic picture, but rather to remind readers that numerical results must be treated with an appropriate level of caution.

One way around this predicament is to incorporate realistic microphysics into MHD codes in some approximate manner. Heuristic, parameterized transport coefficients that account for non-classical effects are one possibility. Several studies have already approximated the anomalous resistivity due to current-driven instabilities by making the resistivity depend on the ratio of the drift velocity of the electric current to the thermal velocity of the particles (e.g., Ugai, 1999; Yokoyama and Shibata, 2001; Uzdensky, 2003). Hoshino (1991) proposed a different form based on lower hybrid turbulence that depends on $(\nabla B^2)^2$. Heat flux saturation is another example where non-classical microphysics can be incorporated into an MHD or hydro code (see Section 4.2). Further efforts along these lines are strongly encouraged.

### 3.3. SWITCH-ON PROPERTY

We now return to Equation (1) to discuss a fundamental property of DC energy conversion. In our earlier rough estimate of the Poynting flux through the coronal base, we casually assumed that the horizontal and vertical components of the field have comparable magnitudes. Let us instead ask what value of the horizontal component is necessary to make the Poynting flux equal to the observed coronal loss rate. In Parker's picture, coronal flux tubes become tangled by random footpoint motions associated with turbulent convection. As long as the random walk step size is longer than the characteristic separation of nearby tubes, each displacement will bend a flux tube around its neighbor, and the lower part of the tube will always trail behind the moving footpoint[1]. In that case, $\boldsymbol{B_h}$ and $\boldsymbol{V_h}$ will be approximately aligned, so that $|\boldsymbol{B_h} \cdot \boldsymbol{V_h}| \approx B_h V_h$. Using typical observed values for $F$, $B_v$, and $V_h$ in the quiet Sun, Equation (1) implies that the field is tilted from vertical by $\arctan(B_h/B_v) \approx 20^o$. Active region values give an angle closer to $10^o$, while Parker (1988) estimated an angle roughly in between these two. We do not wish to imply that these determinations are highly accurate, but it is clear that the tilt angle is neither very close to $0^o$ nor very close to $90^o$.

This deceptively simple result has a profound implication that has been largely overlooked. Whatever the mechanism of DC coronal heating, it must remain dormant long enough for magnetic stresses to build to the required level, i.e., long enough for the field to become sufficiently tilted, and then it must switch on suddenly to prevent the stresses from becoming too large. If the mechanism were to activate too early or too late, the corona would be cooler or hotter than observed. It is interesting that quiet Sun and active region observations imply roughly the same critical angle. This suggests a common heating mechanism. In Section 7, we equate the $20^o$ tilt angle with a $40^o$ misalignment angle between adjacent flux tubes in the tangled field. We show that the secondary instability of electric current sheets switches on at approximately this value.

## 4. Plasma Response

We now discuss how the plasma responds to heating produced by the dissipation of magnetic stresses or waves. This is extremely important for two reasons. First, the radiative signatures that are used to test any coronal heating theory depend critically on the state of the plasma. Second, the response of the plasma can in some cases greatly affect the subsequent heating.

### 4.1. CONNECTION TO THE LOWER ATMOSPHERE

A fundamental aspect of the plasma response is the close thermal and dynamic connection between the corona and lower atmosphere. This is true for both equilibrium and time-dependent conditions. In the case of static equilibrium, thermal conduction transports somewhat more than half of the coronal heating energy down to the transition region, where it is more efficiently radiated at the higher densities and cooler temperatures (e.g., Vesecky, Antiochos, and Underwood, 1979). Some studies have erroneously treated the corona in isolation, using rather arbitrary heat flux boundary conditions at a location somewhere above the transition region. This is extremely dangerous. Many coronal solutions to the equilibrium equations are unphysical in the sense that there does not exist a matching solution for the transition region. The pressure and heat flux must be the same at the top of the transition region and bottom of the corona, where the two solutions meet. The problem is that, in equilibrium, the pressure of the transition region is determined by the heat flux, and this pressure may be different from the coronal pressure, creating an inconsistency. It is vitally important that the corona and transition region be treated together as a *coupled system*.

---

[1] More precisely, the random walk step size must be longer than the correlation length of the flow, since tubes moving together in a group can wrap around other groups but not each other.

The critical role of the lower atmosphere is even more apparent when the heating is time dependent. An increase in the heating rate causes the coronal temperature to rise, producing an increase of the downward heat flux. The transition region is unable to radiate the additional energy, so heated plasma flows into the corona through the process known as "chromospheric evaporation." The upflow can be quite explosive and even produce shocks if the change in the heating rate is sufficiently intense and abrupt. If the heating rate then decreases, an inverse-like process occurs in which plasma drains from the loop and "condenses" back onto the chromosphere. This is generally more gradual and proceeds on the slower radiative cooling timescale.

Because the plasma and magnetic field are frozen together, and because cross-field thermal conduction is greatly inhibited, it is customary to use 1D hydro simulations to study the plasma response to variations in coronal heating. Simulations of this type, often referred to as "loop models," assume that the magnetic field is rigid and plays only a passive role by channelling the plasma and thermal energy along the field lines. The advantage of this approach is that highly complex field-aligned behavior can be accurately simulated using a full energy equation. The disadvantage is that the heating must be specified and cannot be computed self-consistently. Skeptics have complained that 1D hydro models are of limited value since, for DC heating, the energy conversion necessarily involves changes in the magnetic field that are not included. This criticism is largely over-exaggerated. One can think of the magnetic field as being comprised of many small flux strands, or mini loops, each one thinner than the kilogauss photospheric flux tubes discussed earlier. The field is indeed modified as magnetic energy is converted into heat, but once a new flux strand is formed---from reconnection, for example---its subsequent motion and change in shape have relatively little impact on the evolution of the plasma contained within. The plasma evolution is also independent of how the plasma is evolving in neighboring strands. 1D hydro models are therefore entirely appropriate for investigating the plasma behavior of individual strands.

There is one important caveat, however. This claim is valid only for strands that maintain their integrity for periods longer than a cooling time. Depending on the situation, additional reconnections at other locations along the strand may cause it to exchange sections with several different partners, thereby mixing the plasmas and impacting how the plasma evolves. Another key point is that an observed coronal structure, such as an X-ray or EUV loop, is comprised of many strands that may be heated very differently or at very different times. In this case, each strand must be simulated separately. Computing a single loop model using the average heating for the different strands will produce an entirely incorrect result.

Eventually, 1D hydro models will be completely supplanted by 3D MHD models that accurately treat both the energy conversion and the plasma response. However, until the 3D models incorporate field-aligned thermal conduction and properly resolve the transition region, 1D models will continue to play an important role. We cannot emphasize strongly enough that both the temperature and density of the corona are greatly affected by the heat flux into and through the transition region. Furthermore, the transition region is an important layer to study in its own right. It radiates more than half of the energy deposited in the corona, making it a useful diagnostic of coronal heating (Antiochos *et al*., 2000) and an important source of UV input to the terrestrial upper atmosphere (Lean, 1997).

Some existing MHD models do include field-aligned thermal conduction (e.g., Lionello, Linker, and Mikić, 2001; Yokoyama and Shibata, 2001; Gudiksen and Nordlund, 2002, 2005a,b), but they do not yet adequately resolve the transition region. To understand the resolution requirements, we can estimate the temperature scale length at temperature $T$ in the transition region to be

$$H_T \approx \frac{7}{2}\left(\frac{T}{T_m}\right)^{7/2} L \ , \tag{3}$$

where $T_m$ and $L$ are the maximum coronal temperature and half length of the loop strand (more precisely, $L$ is the distance along the strand between the chromosphere and the position of maximum temperature). This simple result is derived using $H_T = T/|dT/ds|$, $F_c = -\kappa_0 T^{5/2} dT/ds = -(2/7)\kappa_0 d(T^{7/2})/ds$, and assuming that the heat flux is roughly constant in the middle and upper transition region and given by $F_c \approx -(2/7)\kappa_0 T_m^{7/2}/L$. For $T/T_m = 0.1$ and $L = 5 \times 10^9$ cm, Equation (3) gives $H_T \approx 5 \times 10^6$ cm. We have verified that this is a reasonable approximation for an exact equilibrium solution with $T_m = 2$ MK (T = 0.2 MK). To resolve this temperature scale length requires a mesh spacing of 20 kilometers or less. The requirements are even more severe at greater depths in the transition region and for hotter and shorter strands. Not resolving the transition region can produce errors of a factor of 2 or more in the coronal density and 4 or more in the predicted radiation (S. Patsourakos, private communication).

The difficulties are compounded when the heating is variable, because then the transition region moves up and down the strand in response to the changing coronal pressure. For highly impulsive heating, the vertical displacement is approximately

$$\Delta z \approx -H_g \ln\left(\frac{2}{3}\frac{e}{P}\right) \ , \tag{4}$$

where $H_g$ is the gravitational scale height of the chromosphere, $e$ is the strand-averaged energy per unit volume of the heating event, and $P$ is the pre-event pressure at the top of the chromosphere. In the simulation we show below, the transition region moves by approximately 4 scale heights, or $7 \times 10^8$ cm for our model chromospheric temperature of $3 \times 10^4$ K. It is clear that adaptive mesh refinement is required to properly model dynamic conditions. To further emphasize this point, we note that moving transition regions can form at locations other than the base of the loop, such as at the sides of coronal condensations (e.g, Karpen *et al*., 2001; Karpen, Antiochos, and Klimchuk, 2005; Mueller, Hansteen, and Peter, 2003). Propagating shocks may also be present in many situations.

## 4.2. Impulsive Heating

As discussed above, it is imperative to treat individual magnetic flux strands in order to understand how the plasma responds to heating. We define a strand to be a mini-loop for which the heating and plasma properties are approximately uniform on a cross section. From this perspective, we find that most coronal heating theories predict that the heating is impulsive in nature. Petschek-type magnetic reconnection is a good example. Individual strands are heated only while they are in contact with standing slow shocks that emanate from the reconnection X-point. A newly reconnected strand snaps away from the X-point at the Alfvén speed, so the heat pulse in the strand lasts a few tens of seconds at most (e.g., a $2 \times 10^9$ cm long shock and $2 \times 10^8$ cm s$^{-1}$ Alfvén speed give a pulse of 10 s). Even if the reconnection is steady and persists unchanged for long periods, it is the heating of individual strands that matters, and this is very short lived. Resonant wave absorption is another example of heating that is usually thought of as steady, but in fact is highly impulsive when looked at from the proper perspective (see below).

Loop models with impulsive heating are therefore especially relevant (e.g., Reale *et al.*, 2000; Warren, Winebarger, and Hamilton, 2002; Klimchuk, 2002). We now discuss the results of one such simulation in some detail (Klimchuk, Patsourakos, and Winebarger, 2003). It was performed with our state-of-the-art hydro code, ARGOS, that uses adaptive mesh refinement (Antiochos *et al.* 1999). An important new feature is the inclusion of heat flux saturation. The model strand consists of a semi-circular coronal section of length $1.5 \times 10^{10}$ cm with long chromospheric sections attached to both ends. We start with an initial static equilibrium that is maintained at a peak temperature of only 0.5 MK by a weak uniform heating of $1.0 \times 10^{-6}$ erg cm$^{-3}$ s$^{-1}$. We then rapidly increase and decrease the heating by a factor of 1500 using a triangular profile lasting 500 s. The heat pulse is spatially uniform and corresponds to a $2 \times 10^{24}$ erg nanoflare for a strand cross-sectional radius of $10^7$ cm.

Figure 2 shows the resulting evolution of temperature, density, and velocity averaged over the central 80% of the coronal section. The generic response to impulsive heating is well known (e.g., Cargill 1994). The spatially-averaged temperature rises rapidly to a maximum of about 8 MK near the peak of the energy release. An intense heat flux drives strong chromospheric evaporation with upflow velocities reaching $3 \times 10^7$ cm s$^{-1}$. The initial cooling is quite rapid and is dominated by thermal conduction. Material continues to evaporate into the loop during this phase. As temperature falls and density rises, radiation becomes progressively more important. Eventually, after about 1500 s, it takes over as the primary cooling mechanism. The strand then enters a long period of relatively slow cooling and mild condensation, when temperature and density decrease together. It is during this second phase that most of the emission is produced. The small wiggles in the curves of Figure 2 are due to waves that are excited by the impulsive heating and propagate back and forth along the strand with diminishing amplitude.

Despite the strong evaporation, this model does not predict that easily detected blue-shifted emission would be observed in most spectral lines. The reason is that the highest speed upflows are both short lived and faint. They have very high temperatures and very low densities and emission measures. For most spectral lines, the temporally averaged emission from the strand (or the spatially averaged emission from a bundle of randomly heated strands) is dominated by the much slower draining phase. Only small Doppler shifts are predicted, consistent with observations (Warren and Doschek, 2005; Patsourakos and Klimchuk, 2005b). Very hot emission lines are an exception. They are relatively insensitive to the low temperatures of the draining phase, and if they are "tuned" to the temperature of the explosive upflow, they can exhibit distinctive blue-wing enhancements (Patsourakos and Klimchuk, 2005b). These features may be detectable by future spectrometers like the one that will fly on *Solar-B*, though the lines will be quite faint and long integration times may be required.

Figure 3 shows the instantaneous state of the strand at time $t = 220$ s, roughly halfway into the energy release. Temperature and density are plotted as a function of position, $s$, along the strand. The top of the chromosphere is located at positions $s = 5 \times 10^9$ and $2.2 \times 10^{10}$ cm in the left and right legs, respectively. These positions are depressed relative to the initial positions, as discussed in the previous section. The solid and dashed curves represent two separate simulations that differ only in the way thermal conduction is treated. For the dashed curves, the well-known Spitzer-Harm formula for classical thermal conduction is used at all times:

$$F_c = -\kappa_0 T^{5/2} \frac{\partial T}{\partial s} \quad , \tag{5}$$

where $\kappa_0 = 10^{-6}$ in cgs units. This formula is appropriate under most conditions. However, early in the simulation, when the densities are low and the temperatures are high, it predicts a greater heat flux than is physically possible. The real heat flux saturates at a magnitude given by

$$F_s = c \frac{3}{2} \frac{k^{3/2}}{m_e^{1/2}} n T^{3/2} \quad , \tag{6}$$

where $n$ is the electron number density, $k$ is Boltzmann's constant, $m_e$ is the electron mass, and $c$ is a flux limiter constant that we set to 1/6 (Luciani, Mora, and Virmont 1983; Karpen and DeVore 1987). The simulation represented by the solid curves properly accounts for heat flux saturation. This is the same simulation shown in Figure 2. The plasma response is strikingly different in the two simulations. Much higher temperatures are reached when the heat flux is saturated because more of the energy is bottled up in the central portion of the strand. These large differences do not persist for long, however. Only the hottest spectral lines are expected to be significantly affected, but these are precisely the lines that are most useful for diagnosing impulsive heating.

Thus far, we have resisted using the term "nanoflare" to describe impulsive heating in a loop strand, because many people think of a nanoflare as an identifiable event in a distinguishable solar feature, such as a transient X-ray brightening (e.g., Shimizu, 1995). Such an event would require many strands of the type we have discussed. In the subsequent discussion, a nanoflare refers simply to an impulsive energy release in a single, probably unresolvable, strand.

A very important variable is the frequency with which nanoflares recur in a given strand. The properties of the plasma depend sensitively on whether the time interval between successive events is long or short compared to a cooling time (Kopp and Poletto, 1993; Walsh, Bell, and Hood, 1997; Mendoza-Briceno, Erdélyi, and Sigalotti, 2002; Testa, Peres, and Reale, 2005). The greater the time lag, the more the strand deviates from equilibrium. When nanoflares repeat very rapidly, the situation is effectively the same as steady heating, and the strand persists in a quasi-equilibrium state. Another important variable is the rate at which individual nanoflares turn on and off (Reale *et al.*, 2000; Spadaro *et al.*, 2003). The more slowly the heating rate changes, the more closely the behavior resembles quasi-static evolution.

4.3. RESONANT WAVE ABSORPTION

Most heating mechanisms are not significantly affected by the plasma response, but resonant wave absorption is one important exception where a critical feedback occurs. As pointed out in Section 2.2, substantial wave energy can enter through the base of a loop only at discrete frequencies. These are the natural resonance frequencies of the loop, and the oscillations that are excited are called global modes, or collective modes, because all of the magnetic strands oscillate together in a kink-like fashion. Despite oscillating in unison, each strand has a preferred resonance frequency that depends on the local density through the Alfvén speed:

$$\omega_{res} \propto \frac{V_A}{L} \propto n^{-1/2} \ , \tag{7}$$

where we have made use of the fact that the magnetic field strength is nearly constant over the cross section of a low-$\beta$ loop. If density varies over the cross section, only some of the strands will be in tune with the global oscillation. Energy from the global motions is preferentially pumped into these strands, and they experience large-amplitude torsional oscillations. Steep velocity and magnetic gradients develop between these and nearby out-of-tune strands, and the plasma is heated in narrow resonance layers.

Nearly all of the theoretical work on resonant wave absorption begins with an assumed density profile, $n(r)$, describing the variation of density with radial distance away from the loop axis. A typical example is shown by the dashed curve in Figure 4a, where the density peaks at the axis and monotonically decreases to an external value. This density profile is assumed to be constant in time. Driver motions are applied, either at or near the global mode frequency or with a broad frequency spectrum. Collective oscillations are set up at the fundamental frequency and perhaps also some harmonics. One or more resonance layers develop at fixed radial positions, and the loop reaches a steady state with constant, non-uniform heating (e.g., Poedts, Goossens, and Kerner, 1989; Steinolfson and Davila, 1993; Ofman, Davila, and Steinolfson, 1994; Erdélyi and Goossens, 1995; De Groof and Goossens, 2002). The solid curve in Figure 4a is a schematic representation of the heating profile corresponding to a single resonance layer.

Unfortunately, there is a fundamental inconsistency with this picture. We have already seen that strong heating leads to large coronal densities through evaporation, while weak heating leads to small coronal densities through condensation. The original density profile will not remain constant, as assumed. Instead, it will evolve to a profile roughly resembling the heating profile, as represented by the dashed curve in Figure 4b (artificially broadened for clarity). This will change the resonance conditions, and different strands will come in tune with the global oscillations. The old resonance layer will shut off and two new ones will activate, as indicated by the solid curve in Figure 4b. This new heating profile will lead to a new density profile, which will in turn lead to a new heating profile, etc. We quickly see that resonant wave heating is an inherently dynamic process. No steady equilibrium is possible.

We have verified this scenario with an MHD simulation in a slab geometry that mimics a coronal loop (Ofman, Klimchuk, and Davila, 1998). Chromospheric evaporation and condensation are treated only approximately using a quasi-static equilibrium scaling law to update the density in response to the computed heating:

$$n(r) \propto Q(r)^{5/7} \ , \tag{8}$$

where $Q(r)$ is the volumetric heating rate. Instead of one stationary resonance layer, there are multiple layers that drift back and forth throughout the loop. A key point is that, even though the spatially averaged heating rate is approximately constant, the different strands that make up the loop are heated *impulsively* as resonance layers pass by. This is one more example, like Petschek reconnection, where seemingly steady heating is actually impulsive when looked at from the proper perspective. Future work on resonant wave heating must incorporate the important feedback between the energy release and plasma response.

We quickly remark that the global loop oscillations discussed above are primarily of a kink nature and are excited by translational driver motions. Energy can transfer into the torsional oscillations at the resonance layers because the global modes have both kink and torsional components (e.g., Davila, 1987). Torsional driver motions are also possible, but they tend to produce heating by the different process of phase mixing. In that case, the different cylindrical shells of the loop oscillate independently of each other and extract power from the driver at different preferred frequencies. There is no collective behavior. Steep gradients develop even if adjacent shells oscillate with similar amplitude, because the oscillations are out of phase.

## 5. Radiation

The next step in the coronal heating flowchart is to determine the radiation spectrum emitted by the heated plasma. This is relatively straightforward if the plasma is in a state of ionization equilibrium, because then powerful atomic physics software packages such as CHIANTI (Dere *et al*. 1997) can be used. The emissivity of the plasma in a particular spectral line is given by $n_e^2 G(T)$, where $G(T)$ depends on atomic physics parameters and is unique to each spectral line. Since the coronal plasma is optically thin, one simply integrates the emissivity along the line-of-sight to determine the intensity of the radiation. Synthetic spectral line profiles can be constructed by assuming a Gaussian at each line-of-sight position that is broadened according to the local temperature and Doppler shifted according to the local velocity. Uncertainties are quite large, however, due to uncertainties in the ionization rates, recombination rates, etc. that go into $G(T)$. Factors of two are not uncommon (Savin and Laming 2002). Elemental abundances are also problematic. Absolute abundances are difficult to measure, and relative abundances are known to vary with time and from one solar feature to the next (e.g., Feldman and Widing, 2003).

It is much more complicated to determine the radiation spectrum when the plasma is not in ionization equilibrium. This is likely to be the case if there is rapid evolution due to impulsive heating or rapid cooling (e.g., Raymond, 1990), or if there is a flow through a steep temperature gradient (e.g., Roussel-Dupree, 1980). The determining factor is whether the timescale for temperature change in a plasma parcel is short or long compared to the ionization or recombination time, whichever is relevant. Consider an observation of the simulation in Figure 3 with the 193 Å line of Ca XVII, which is formed at 5 MK under equilibrium conditions. This is one of the lines that will be observed by the Extreme-ultraviolet Imaging Spectrometer (EIS) on *Solar-B*. Because the densities are so low at this early stage of the simulation, the ionization equilibration time at the midpoint of the strand is roughly $10^4$ s (Mazzotta *et al.*, 1998). Temperature is increasing with a much shorter timescale, so the ionization lags behind.

In situations like this when ionization equilibrium does not apply, it is necessary to solve the ionization rate equations in order to accurately determine the radiation spectrum (e.g., Mariska *et al.*, 1982; Spadaro, Leto, and Antiochos, 1994; Sarro *et al.*, 1999; Bradshaw and Mason, 2003a,b; Mueller, Hansteen, and Peter, 2003). Fortunately, this can usually be done as a secondary step, after the hydrodynamic or MHD simulation has been completed. The reason is that nonequilibrium effects tend to have minimal impact on the spectrum-integrated radiation loss rate and therefore do not significantly modify the plasma evolution (although see Spadaro *et al.*, 1990).

Other microphysics effects that can alter the radiation signature include the thermal force, which is important in regions of steep temperature gradient (e.g., Woods and Holzer, 1991), and ambipolar diffusion of neutral atoms relative to charged species, which is important at interfaces with chromospheric plasma, either at the base of loops or at the boundaries of prominences and coronal condensations (e.g., Fontenla, Avrett, and Loeser, 1990, 1991). Non-local heat transport associated with electrons in the high-energy tail of the particle distribution can also be important (e.g., Karpen and DeVore, 1987). Scudder (1992) has even proposed that the coronal plasma is essentially derived from suprathermal particles from the transition region. If this were the case, however, we would not expect the corona to be so different in the quiet Sun and coronal holes, since the transition region is quite similar in these regions (S. Antiochos, private communication).

## 6. Observables

The last step in the coronal heating flowchart is to predict observables that can be compared directly with real solar data. As sophisticated as solar instruments are, they only detect bits and pieces of the emitted spectrum, and often with a large degree of averaging over space, time, and wavelength. This can be a great source of ambiguity and confusion, and care must be taken not to misinterpret the observations or to present legitimate interpretations as unique when in fact they are not. The ambiguities are most apparent when one realizes that the plasma observed in a single observational pixel may be highly nonuniform in both temperature and density. Sub-resolution structures are both theoretically predicted and observationally inferred (e.g., Orrall *et al.*, 1990; Brosius, Davila, and Thomas, 1996; Schmelz 2002). Furthermore, any line-of-sight path through the optically-thin corona can cross multiple large-scale structures with different properties.

It is common to use inversion methods to infer physical quantities such as temperature, emission measure, density, and filling factor. The approach can be quite useful, even when the observed plasma is nonuniform, but the quantities must be appropriately interpreted as weighted averages (Klimchuk and Cargill, 2001). Problems arise when the spectral information is severely limited, such as when the filter ratio technique is used. The ratio of intensities observed through two filters implies a temperature, but that temperature has clear meaning only if the plasma is known to be isothermal. There are many examples of *TRACE* observations where the addition of a third filter reveals that the plasma is actually multi-thermal with a broad temperature distribution (Noglik, Walsh, and Ireland, 2004, 2005; Patsourakos and Klimchuk, 2005c). The pitfalls of the filter ratio technique have been well documented (Reale and Peres, 2000; Martens, Cirtain, and Schmelz, 2002; Schmelz *et al.*, 2003; Weber *et al.* 2005), although see Aschwanden (2002).

Forward modeling is the safest way to compare theory and observation. With this approach, a simulated observation (e.g., the intensity of a spectral line or in a passband) is derived from a model and compared with real data. The parameters of the model are varied until, hopefully, a satisfactory fit is achieved. Forward modeling *must* be used whenever ionization nonequilibrium applies.

## 7. Tangled Fields, Secondary Instability, and Nanoflares

We now combine some of the ideas we have already introduced and discuss what we believe to be the most promising explanation of coronal heating. We start with Parker's (1983, 1988) picture of tangled magnetic fields. The kilogauss photospheric flux tubes that are randomly shuffled by turbulent convection contain only about $10^{17}$ Mx of flux (e.g., Solanki, 1993; Muller, 1994; Socas-Navarro and Sánchez Almeida 2002). Many tens to hundreds of them must therefore be bundled together within a single coronal loop observed by *TRACE* or *Yohkoh* (e.g., Priest, Heyvaerts, and Title, 2002). These elemental tubes become wrapped and braided by the footpoint motions, and current sheets form at the interfaces where the tubes are misaligned. Parker suggested that energy is released impulsively as nanoflares when the stresses and currents become too large.

As we pointed out in Section 3.3, energy balance considerations indicate that the field at the coronal base is tilted from vertical by approximately $20^{\circ}$. If we associate this with tangling, it equates to a critical misalignment angle of $40^{\circ}$. For many years there was no physical explanation for why nanoflares should switch on at this particular value, but we have recently identified a mechanism called the secondary instability that has precisely this property (Dahlburg, Klimchuk, and Antiochos, 2003, 2005). It is a rapidly growing instability that sets in after the primary instability, the well-known tearing mode, saturates at a low level. As the secondary instability becomes nonlinear, it transitions to turbulence and produces intense, impulsive heating. Figure 5 shows the combined resistive and viscous heating rate as a function of time from one of our 3D MHD simulations. The event lasts about 100 s and would release approximately $10^{24}$--$10^{25}$ ergs of energy in a current sheet of realistic size. Note that this duration and energy are similar to what we assumed for the nanoflare hydro simulation presented in Section 4.2. Impulsive heating of this type is able to explain important properties of many observed coronal loops that cannot be understood with steady heating, as we now explain.

## 7.1. OVER AND UNDER DENSITIES

A large majority of warm ($T \sim 1$ MK) coronal loops observed by *TRACE* and *EIT* are over dense compared to what is expected for static equilibrium (Aschwanden *et al.*, 1999; Aschwanden, Schrijver, and Alexander, 2001; Winebarger, Warren, and Mariska, 2003) or steady flow equilibrium (Patsourakos, Klimchuk, and MacNeice (2004). The discrepancy is reduced, but not eliminated, if the heating is assumed to be concentrated near the loop footpoints. In contrast, hot ($T > 2$ MK) loops observed by *Yohkoh* are under dense compared to static equilibrium (Porter and Klimchuk, 1995). Loops of intermediate temperature observed by the SXI instrument on *GOES-12* have about the right density (López Fuentes, Mandrini, and Klimchuk, 2004). This may indicate three physically distinct classes of loops, perhaps heated in completely different ways, but there is another possibility that unifies the results into a single picture.

The over and under densities are related to the ratio of the radiative to conductive cooling times:

$$\frac{\tau_{rad}}{\tau_{cond}} = 2.9 \times 10^{-7} \frac{T^{7/2}}{n^2 L^2 \Lambda(T)} \quad , \qquad (9)$$

where $\Lambda(T)$ is the optically-thin radiative loss function. Radiation and thermal conduction losses are comparable in equilibrium loops (Vesecky, Antiochos, and Underwood, 1979), and therefore the cooling time ratio should be close to unity for loops that are near equilibrium. Figure 6 shows the actual ratios computed for a collection of 56 different loops observed by either *Yohkoh* or *TRACE* and plotted against the observed temperature (Klimchuk, Patsourakos, and Winebarger, 2003). The temperatures and emission measures were determined using filter ratios, and then the densities were determined using the observed loop widths assuming a circular cross section (Klimchuk, 2000) and a filling factor of one. Notice that the hot loops lie above the horizontal line $\log(\tau_{rad}/\tau_{cond}) = 0$, indicating that thermal conduction dominates over radiation, and the warm loops lie below the line, indicating that radiation dominates over thermal conduction. This is precisely what would be expected if the loops were cooling after having been impulsively heated to very high temperatures (Section 4.2). The points can be lowered by assuming filling factors less than one, so the hot loops can be made to be consistent with static equilibrium, but the warm loops would then become even more discrepant.

The curve in Figure 6 shows the cooling track of a nanoflare simulation similar to the one presented in Figures 2 and 3. In this case, the nanoflare is 7 times more energetic and is repeated every 5000 s. Under these circumstances, the density at the beginning of each event is about $4 \times 10^8$ cm$^{-3}$. The track begins at the top-right, at the time of maximum temperature, and progresses downward and to the left. The agreement between the simulation and observations is remarkably good. One is tempted to conclude that the simulation is representative of all 56 loops, and that the loops look different simply because they are observed at different phases in the cooling. This simple interpretation is not correct, however. *Yohkoh*, *TRACE*, and SXI/*GOES* loops are observed to evolve much more slowly than the simulation would predict, i.e., they are observed to live much longer than a cooling time (Porter and Klimchuk, 1995; Winebarger, Warren, and Seaton, 2003; López Fuentes, Mandrini, and Klimchuk, 2004). The loops cannot be monolithic structures cooling as whole units. Rather, each loop must be a bundle of unresolved strands that are heated at different times. The loop will then appear to evolve slowly even though the individual strands evolve rapidly. *Yohkoh* preferentially detects the hottest strands that are in the early, conduction-dominated phase of cooling, and *TRACE* preferentially detects the warm strands that are in the late, radiation-dominated phase.

The two diamonds in Figure 6 indicate synthetic *Yohkoh* and *TRACE* observations obtained from the time-averaged emission from the entire nanoflare simulation. This is equivalent to an instantaneous snapshot of a bundle of identical strands in random stages of cooling. The diamonds lie well within the band of real data points, demonstrating that our explanation is at least plausible. The fact that the data points form a band, rather than two tight clusters around the diamonds, suggests that real nanoflares have a range of energies. This is not surprising, given the wide range of magnetic field strengths that exist in the corona.

The basic picture of multi-stranded, impulsively heated loops is certainly not new. It was first put forward by Cargill (1994) and has subsequently been studied in considerable detail (e.g., Cargill and Klimchuk 1997, 2004; Vekstein and Katsukawa, 2000; Klimchuk and Cargill 2001; Warren, Winebarger, and Hamilton 2002; Warren, Winebarger, and Mariska 2003, Patsourakos and Klimchuk 2005a,b). The model is extremely appealing in its ability to tie together a number of different observations and concepts. Significant issues remain, however. The biggest concern is that the model predicts that loops should be visible either simultaneously or nearly simultaneously over a broad range of temperatures. In particular, a loop should be detectable by both *Yohkoh* and *TRACE* at the same location, either at the same time or with a slight delay. How often this is actually the case has been actively debated (e.g., Nitta, 2000; Schmelz *et al.*, 2001; Schmelz, 2002; DelZanna and Mason, 2003; Nagata *et al.*, 2003; Landi and Landini, 2004; Schmieder *et al.*, 2004; Aschwanden and Nightingale, 2005; Aschwanden, 2005; Reale and Ciaravella, 2005; Schmelz and Martens, 2005). Some have argued that hot and warm loops occur preferentially in different parts of active regions and therefore tend to be mutually exclusive. Others suggest that hot and warm loops are located near each other, but do not actually overlap. Still others have presented examples of loops that can be clearly seen in a variety of spectral lines, though with a spatial resolution that is less than ideal.

Cases have recently been reported where a loop is first seen by *Yohkoh* and later seen by *TRACE*, sometimes with a temporal overlap (Winebarger and Warren, 2005; Ugarte-Urra, Winebarger, and Warren, 2005). This suggests that a "storm" of nanoflares has occurred over a limited time window. We emphasize that a single large nanoflare can be ruled out, because the loops live considerably longer than a cooling time, irrespective of the observing instrument. Preliminary indications are that the delay between when the loop is first seen by *Yohkoh* and first seen by *TRACE* is longer than predicted by cooling strand models (~2000 and 3000 s in the simulations presented here). Further study is necessary to determine just how serious this problem is.

The issue of cospatial hot and warm loops can be avoided if nanoflares recur frequently *within each strand*. Repetition times much shorter than a cooling time produce quasi-static equilibrium conditions. Larger events and higher frequencies maintain hot strands, while smaller events and lower frequencies maintain warm strands. The hot and warm strands need not be intermingled. If all of the nanoflares within a loop bundle have a similar magnitude and recur with a similar frequency, then the loop will be approximately isothermal. The drawback of this picture is that it is unable to explain the observed under densities of *Yohkoh* loops and over densities of *TRACE* loops. The under densities can be corrected by assuming a small filling factor (Porter and Klimchuk, 1995), but the over densities present a fundamental observational challenge.

Regardless of the temporal nature of the heating, we can ask how thick the basic isothermal strands are. This provides vital information on the cross-field scale of the energy release. Note that the concept of an isothermal strand is just a convenience, since temperature most likely has a continuous variation across the magnetic field. We define a strand to be a flux tube that is approximately isothermal over a cross section. Aschwanden and Nightingale (2005) and Aschwanden (2005) have argued that some of the thinnest loops seen by *TRACE* are isothermal strands. However, only 14% of their thin-loop data could be fitted with an isothermal model to within a rather generous reduced-$\chi^2$ of 1.5 (a reduced-$\chi^2$ fit of 1.5 means there is only a 13% probability that the model is correct). The remaining 86% of the data have even larger reduced-$\chi^2$ values and are clearly inconsistent with the isothermal model. We conclude that a large majority of strands are thinner than and perhaps much thinner than $2 \times 10^8$ cm. Whether the strands that are seen together as a multi-thermal loop are closely spaced and physically related or simply a chance alignment of structures that are widely separated along the line-of-sight is currently being investigated.

A majority of coronal emission comes from a diffuse component rather than distinct coronal loops. *TRACE* loops, for example, are only about 10-20% brighter than the background. Is the diffuse emission produced by steady heating, as proposed by Antiochos *et al.* (2003), or is it also produced by unresolved, impulsively-heated strands? We have investigated the diffuse emission from one active region observed by *TRACE* and find that it is consistent with the latter picture. The intensity of moss regions, corresponding to the opposite polarity footpoints of a magnetic arcade, and the intensity of the inter-moss region, corresponding to the top of the arcade, have the correct relative magnitudes for impulsively-heated strands. This is suggestive but not definitive. The long-ignored diffuse corona is very important and deserves much more attention in the future.

## 7.2. CORONAL RECONNECTION

The premise that turbulent convection shuffles the footpoints of coronal flux tubes leads us to important conclusions about coronal heating. These conclusions are based solely on the evolution of the magnetic field and do not depend on the plasma response or the requirement that heating produce the observed coronal densities, temperatures, etc. The first conclusion is that magnetic reconnection must be occurring at significant altitudes in the corona, above the magnetic carpet. Reconnection is here broadly defined to be any process that changes the connectivity of the field, including the secondary instability. The justification for our conclusion is straightforward. Since random footpoint motions act to increase the tangling of the field, some other process must act to decrease it. Otherwise, the tangling would grow indefinitely, magnetic stresses would build to ridiculous levels, and no coherent structures such as coronal loops would exist.

Reconnection is the only process that can disentangle the field[2]. It unhooks flux tubes that have become entwined. A crucial point is that both of the reconnecting flux tubes must extend above the magnetic carpet (above about $5 \times 10^8$ cm; Close *et al*., 2003). Reconnection between a long coronal flux tube and a short carpet loop actually increases the coronal tangling (Schrijver *et al*. 1998). Such a reconnection effectively causes the footpoint of the long tube to jump discontinuously from one end of the short loop to the other. Since the carpet fields are disorganized, this produces a random walk in much the same way that turbulent convection produces a random walk. Reconnection involving carpet fields may be an important source of heating in the chromosphere (Schrijver, 2005), in low-lying cool ($T < 0.2$ MK) loops (Antiochos and Noci, 1986; Feldman, Widing, and Warren, 1999, 2000; Peter, 2001), and in the lower legs of warm and hot loops (Aschwanden, Schrijver, and Alexander, 2001), but higher altitude reconnection must play an important, if not dominant, role in heating the corona. This statement is especially true of active regions, where little mixed polarity field exists and the carpet is greatly diminished.

Another conclusion we can draw is that the reconnection process that disentangles the field is likely to include a significant amount of resistive dissipation. This is because footpoint shuffling increases the helicity of the field together with the tangling. According to Taylor's (1974) well-established hypothesis, processes that are essentially ideal (non-resistive) can eliminate the tangling, but not the helicity. When two entwined flux tubes reconnect by Petschek reconnection, the mutual helicity of their interlinking is converted into self helicity of internal twist. This twist helicity must eventually be eliminated. In principle, two tubes with opposite twist can interact and their helicities will cancel. However, tubes with opposite twist and like polarity (having axial fields pointed in the same general direction) are not prone to interact (Linton, Dahlburg, and Antiochos, 2001; case RL0 in their notation). Tubes with opposite twist and opposite polarity (case RL4) can more readily interact, but they are not expected to come into contact very often. Helicity can also be removed from the corona by the submergence or ejection of flux. Submergence may be important in the magnetic carpet, and ejection may be important at the peripheries of streamers or in major events like coronal mass ejections, but whether these processes are important for most coronal flux tubes is unclear. It is possible that much of the helicity introduced by random footpoint shuffling is eliminated by a process that involves substantial resistive dissipation (S. Antiochos, private communication; although see Berger, 1984). Taylor's hypothesis does not apply in that case. The secondary instability is especially appealing in this regard because strong resistive dissipation is expected to occur when the instability transitions to a highly turbulent state.

## 7.3. CORONAL LOOP SYMMETRY

The tangling of elemental flux tubes may have important implications for the overall geometry of coronal loops in addition to their internal structure (López Fuentes, Klimchuk, and Démoulin, 2005). Coronal loops are observed to be symmetric in the sense that the two legs have comparable thickness. The symmetry is much greater than predicted by standard magnetic field extrapolation models. However, due to limitations in the models and the magnetograms on which they are based, the extrapolations do not include the small-scale structure associated with magnetic field tangling. We suggest that this is a critical missing ingredient and propose that a combination of footpoint shuffling and coronal reconnection can explain the observed loop symmetry.

---

[2] Some random walk steps will temporarily disentangle the field, but the net effect of multiple steps is to increase the level of tangling.

To justify our conjecture, we first point out that asymmetric loops are a natural consequence of organized photospheric shear flows. Figure 7a is a simple example of a sheared arcade in which all of the field lines lie in vertical planes at 45° angles to the neutral line (dashed). The flow field that produces this arcade has the form $V_y \propto x$, where $x=0$ is the neutral line, as is indicated by the velocity vectors at the bottom of the figure. We can define a loop (a large flux tube) within this arcade by specifying a circular cross section at the photosphere on the left side of the neutral line. This specification is made *after* the flow has sheared the field. The field lines originating in the circle define the loop volume. They map to a distorted oval on the opposite side of the neutral line. It is clear that the cross section is highly variable along the loop and that the loop would appear highly asymmetric when viewed from most angles[3]. Note that the many elemental flux tubes that we imagine to be contained in the loop are perfectly aligned with their neighbors.

Now suppose that small-scale random motions are superposed on the large-scale shear flow and that the random motions have greater velocity, as typically observed. Starting with the unsheared arcade, it is clear that the field will become progressively more tangled as its shear is increased. The tangling can only proceed so far, however, before the secondary instability switches on and causes elemental flux tubes to reconnect with their neighbors. We expect that a fundamental consequence of these reconnections is that no two flux tubes can have footpoints that are close together at one end, but far apart at the other end. Such a situation would require a high degree of tangling and very large misalignment angles (the misalignment angle scales with the path length of the random walk in the photosphere). If we are correct, then a compact loop cross section on one side must map to a compact loop cross section on the other side. In particular, a circle cannot map to a highly distorted oval as in Figure 7a, but instead must map to an irregular patch as in Figure 7b. The resulting loop would be roughly symmetric.

A key point is that magnetic reconnection changes the connectivity of the field. Magnetic footpoints, as would be viewed in a magnetogram movie, continue to wander farther and farther apart as time progresses, but the way in which these footpoints are connected to each other is altered with each reconnection event. Consider the three elemental flux tubes (field lines) in Figure 7. Footpoints A, B, and C map to $A_1$, $B_1$, and $C_1$ in case (a), without random shuffling, but they map to $A_2$, $B_2$, and $C_2$ in case (b), with random shuffling. The flux tubes in (b) have undergone multiple reconnections with other flux tubes not shown. They are drawn as wavy lines to indicate that the field is tangled. The different locations of $A_2$, $B_2$, and $C_2$ relative to $A_1$, $B_1$, and $C_1$ are due primarily to these reconnections and not (directly) to the footpoint shuffling.

## 7.4. SCALING LAWS

We end this section by mentioning a series of coronal heating studies that make use of theoretical scaling laws that relate either the volumetric heating rate or the input energy flux to parameters such as the field strength and length of coronal flux tubes. Different heating theories make different predictions about the form of the scaling (Mandrini, Démoulin, and Klimchuk, 2000). By determining which scaling best reproduces the observations, it is possible to test the theories. This approach has been applied to individual coronal loops (Klimchuk and Porter, 1995; Mandrini, Démoulin, and Klimchuk, 2000), active regions (Fisher *et al.*, 1998; Démoulin *et al.*, 2003; Yashiro and Shibata, 2001; Lundquist *et al.*, 2004; Mok *et al.*, 2005; Warren and Winebarger, 2005), streamers (Foley *et al.*, 2002), the global Sun (Schrijver *et al.*, 2004), and even other stars (Schrijver and Aschwanden, 2002; Pevtsov *et al.*, 2003). The results favor DC heating over AC heating, but they are unable to clearly discriminate among several competing DC theories. The picture of tangled magnetic fields and a critical misalignment angle (model 2 in Mandrini *et al.*) predicts a volumetric heating rate that varies as $B^2L^{-1}V_h$ and has mixed support. A different version of tangled field heating (Parker, 1983; models 4 and 5 in Mandrini *et al.*) might better reproduce some of the observations, but the heating rate depends on the flux tube diameter, which is typically ignored in the studies even though the diameter may depend on $L$ or $B$ and therefore change the predicted scaling.

---

[3] Had we specified the circular cross section *before* the shear was applied, the final loop would have oval footpoints inclined by 45° angles on both sides of the neutral line, and the loop would be symmetric.

Studies of this type are useful, but it must be recognized that many of the scaling laws are based on plausible conceptual ideas rather than fully developed theories. Often they do not address the specifics of the energy release, including whether the heating is impulsive or steady. A potentially serious shortcoming is that most of the observational comparisons assume that the plasma is in static equilibrium. We have discussed how the observed over density of warm loops is best explained by impulsive heating and nonequilibrium cooling. We suggest that an incorrect assumption of static equilibrium is the reason why models of active regions and the global corona do an especially poor job of reproducing *TRACE* and *EIT* images. Including impulsive heating in these models is an important area of future research.

## 8. Summary and Recommendations

In this article, we have tried to demonstrate how extremely challenging the coronal heating problem is and to emphasize that multiple steps must be accomplished in order to arrive at a definitive solution. In working our way through the coronal heating flowchart, we discovered several important themes. We found that *highly disparate and coupled spatial scales* are involved, both along and across the magnetic field. Small-scale features of particular significance include current sheets, transition regions, resonance layers, and shocks. We also found that physical *connections between the corona and lower atmosphere* are fundamentally important and affect both the source and conversion of energy and the response of the plasma to heating. The properties of the coronal magnetic field, including its energy content, current sheets, and instability, are determined by evolving photospheric boundary conditions. The properties of the coronal plasma, including its temperature, density, and radiation signatures, are greatly influenced by the heat flux to the base and the associated transfer of mass via chromospheric evaporation and condensation. We identified several *microphysics* effects that can play critical roles in the coronal heating problem, such as non-classical particle transport, collisionless reconnection, heat flux saturation, and ionization nonequilibrium. And finally, we showed that most coronal heating mechanisms are impulsive from the perspective of elemental magnetic flux strands, and therefore that *variability and dynamics* are likely to be very important. It is extremely compelling that both MHD simulations of energy conversion and hydrodynamic simulations of plasma response independently point to the same conclusion---that coronal loops are bundles of unresolved, impulsively heated strands.

Many authors use the term "nanoflare" to describe a heating event that, by itself, produces a resolvable feature in a soft X-ray or EUV image. We advocate a much broader definition that includes unresolved strands, and we note that a $10^{24}$ erg energy release---the canonical nanoflare---must occur once every second in a *TRACE* loop in order to satisfy the heating requirements. Studies of isolated point-like brightenings are interesting, but they may have limited relevance for ordinary coronal heating. Furthermore, attempts to estimate the energy content of such events are fraught with uncertainties, and it is not possible to draw firm conclusions about the importance of nanoflares based on extrapolated power-law distributions of the number of events versus event energy (Parnell, 2004).

Where do we go from here? First of all, it is time start bridging the gap between MHD treatments of the energy conversion and hydrodynamic treatments of the plasma response. This recommendation applies both to the way we think about the coronal heating problem and to the way we devise our simulations. One approach that we are planning at the Naval Research Lab is to use the heating rates output from MHD simulations (e.g., Figure 5) as direct input to loop hydro simulations (e.g., Figures 2 and 3). Walsh and Galtier (2000) have already used this approach. This must be done with great care, however, because it cannot be assumed that the average heating rate for the entire MHD domain applies equally to the many flux strands that make up the large-scale magnetic structure (e.g., the loop).

Ultimately, we must solve the full problem with a massive MHD calculation that properly accounts for all of the important physical effects. Such a "grand challenge simulation" is many years off, but it is reasonable to start taking initial steps in this direction. We applaud the ambitious work of Gudiksen and Nordlund (2002, 2005a,b). Although their simulations do not adequately resolve current sheets, transition regions, etc.---and therefore must be treated with great caution--- they are the closest thing yet to a realistic first-principles model of an active region (see also Yokoyama and Shibata, 2001).

Solving the coronal heating problem will also require improved observations. Our lack of discussion on this point should not be taken to imply a lack of importance. Because the corona is structured on very small scales, existing observations are often ambiguous and sometimes misleading (e.g., Section 6). We should not fault the experimenters for this, since it is no less difficult to make a perfect observation than it is to perform a perfect numerical simulation. Trade-offs must always be made when designing new instrumentation. In our opinion, the coronal heating problem does not require observations with large fields of view. An active region size is desirable to allow the observation of entire loops, but significant progress could be made even with a smaller field of view. High temporal resolution and rapid cadence can probably also be sacrificed, if necessary (see below). The greatest emphasis should be placed on high spatial resolution, broad temperature coverage, and high temperature fidelity (ability to discriminate different temperatures). Existing observations lack one or more of these properties. *TRACE* has high spatial resolution, but limited temperature coverage and dubious temperature fidelity. *Yohkoh* has better temperature coverage, but minimal temperature fidelity. The spectrometers on *SOHO* have good temperature coverage and fidelity, but lack spatial and temporal resolution.

There is reason to hope that future imagers, like that proposed for the *Reconnection and Microscale (RAM)* mission, will be able to resolve some of the important small-scale structures that have so far eluded us. For example, if a loop of diameter $d$ contains $N$ elemental strands at different temperatures, the spatial resolution required to discriminate the strands is roughly $d/N^{1/2}$. A resolution of $10^7$ cm may be adequate for strands that correspond to observed kG photospheric flux tubes. However, there would be considerable confusion from line-of-sight overlap effects if many of the strands have similar temperatures. This demonstrates that it may be strategically wise to look for plasma structures that are expected to be the least prevalent (e.g., the hottest strands that presumably persist for the shortest periods of time).

Even if there is reasonable hope of resolving kG flux tubes, we must remember that the small spatial scale of coronal heating is likely to produce elemental strands that are much thinner yet (e.g., at the current sheet interfaces between the kG tubes). Lacking the ability to resolve these strands, we must rely on spectroscopy to sort out the properties of the multi-thermal plasma. A high-speed imaging spectrometer is therefore a top priority. It should observe many emission lines that fully sample the temperature range from about 0.5 MK to 10 MK. The high end of the temperature range is especially important for diagnosing impulsive heating, since relatively little can be learned about the energy release (duration, spatial distribution along the field, etc.) once the plasma enters the slow radiative cooling phase (Winebarger and Warren, 2004, 2005; Patsourakos and Klimchuk, 2005a,b). Since the evolution of individual strands will not be observable if they are unresolved, there is no need for an extremely rapid cadence. An image set every 100 s, or even longer, may be acceptable. The Extreme-ultraviolet Imaging Spectrometer (EIS) that will fly on *Solar-B* has several of the properties we advocate, although its temperature coverage is not as complete as one would like. The proposed *Normal Incidence Extreme UV Imaging Spectrometer (NEXUS)* mission and the *VEry high angular Resolution Imaging Spectrometer (VERIS)* sounding rocket experiment could fill some of the temperature gaps, as could a spectrometer that will presumably fly on *Solar Orbiter*.

Clearly, much work is left to be done, both theoretically and observationally. If we approach the coronal heating problem wisely, we can look forward to a very exciting and productive future!

## Acknowledgements

This work was supported by NASA and the Office of Naval Research. It is a pleasure to acknowledge helpful discussions with many individuals, including, but not limited to, Spiro Antiochos, Peter Cargill, Spiros Patsourakos, Marcelo López Fuentes, Rick DeVore, Mark Linton, Pascal Démoulin, Cristina Mandrini, Dana Longcope, Karel Schrijver, Joe Hollweg, Leon Ofman, Martin Laming, Harry Warren, Amy Winebarger, Joan Schmelz, Markus Aschwanden, Russ Dahlburg, and Judy Karpen. We also thank the referee for a number of constructive comments.

## Figure Captions

*Figure 1.* Flowchart showing the different steps that are involved in solving the coronal heating problem.

*Figure 2.* Spatially-averaged temperature, density, and velocity versus time for a loop strand that has been impulsively heated by a 500 s nanoflare. The averages are computed over the central 80% of the coronal section of the strand. Heat flux saturation is included.

*Figure 3.* Temperature and density versus position along the strand at time $t = 220$ s, roughly halfway into the impulsive energy release. The solid and dashed curves are for simulations with and without heat flux saturation, respectively.

*Figure 4.* Schematic representation of resonant wave absorption. Density (dashed) and heating rate (solid) versus radial distance from the loop axis. Panel (a) corresponds to an initial time, and panel (b) corresponds to a subsequent time, as discussed in the text.

*Figure 5.* Combined resisitive and viscous heating rate versus time for a secondary instability simulation in which the magnetic field rotates by $90^\circ$ across the current sheet (after Dahlburg, Klimchuk, and Antiochos, 2005).

*Figure 6.* Ratio of radiative to conductive cooling times versus temperature for 56 different loops observed by either *Yohkoh* or *TRACE* (crosses). Solid curve shows the cooling track of a nanoflare simulation. Diamonds show synthetic observations based on the time-average of the simulation (*Yohkoh* observation at $\log T = 6.55$; *TRACE* observation at $\log T = 6.09$).

*Figure 7.* View from above of a coronal loop (a large magnetic flux tube) embedded in a translationally symmetric magnetic arcade. The arcade has been sheared by imposing anti-parallel boundary motions (arrows) on either side of the neutral line (dashed). The footpoint cross sections and three field lines, or elemental flux strands, are shown. The loop is defined by the circular cross section on the left, which is specified after the shear is applied. Panel (a) assumes that only the systematic shear flow is present, while panel (b) assumes an additional random component of flow and coronal reconnection.

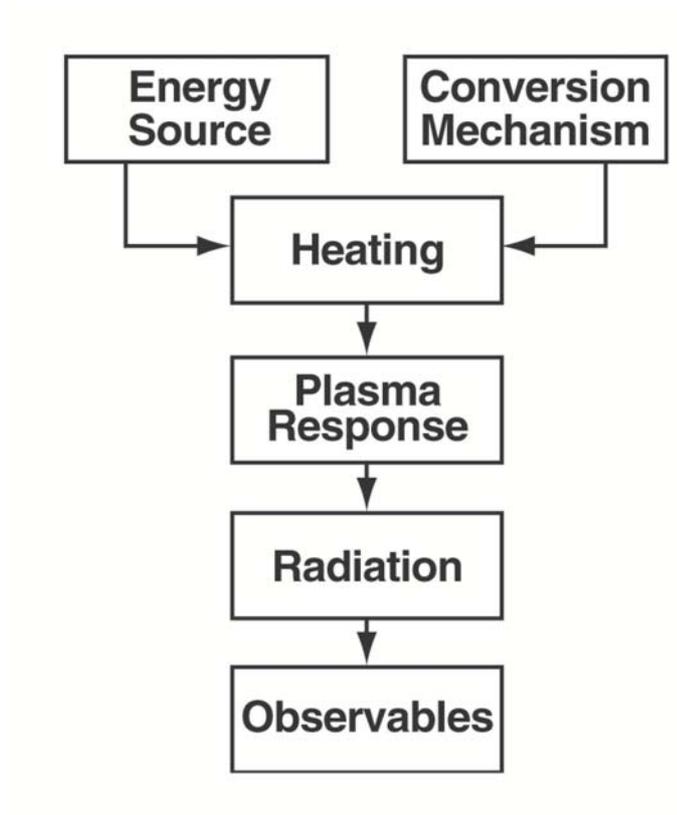

*Figure 1.* Flowchart showing the different steps that are involved in solving the coronal heating problem.

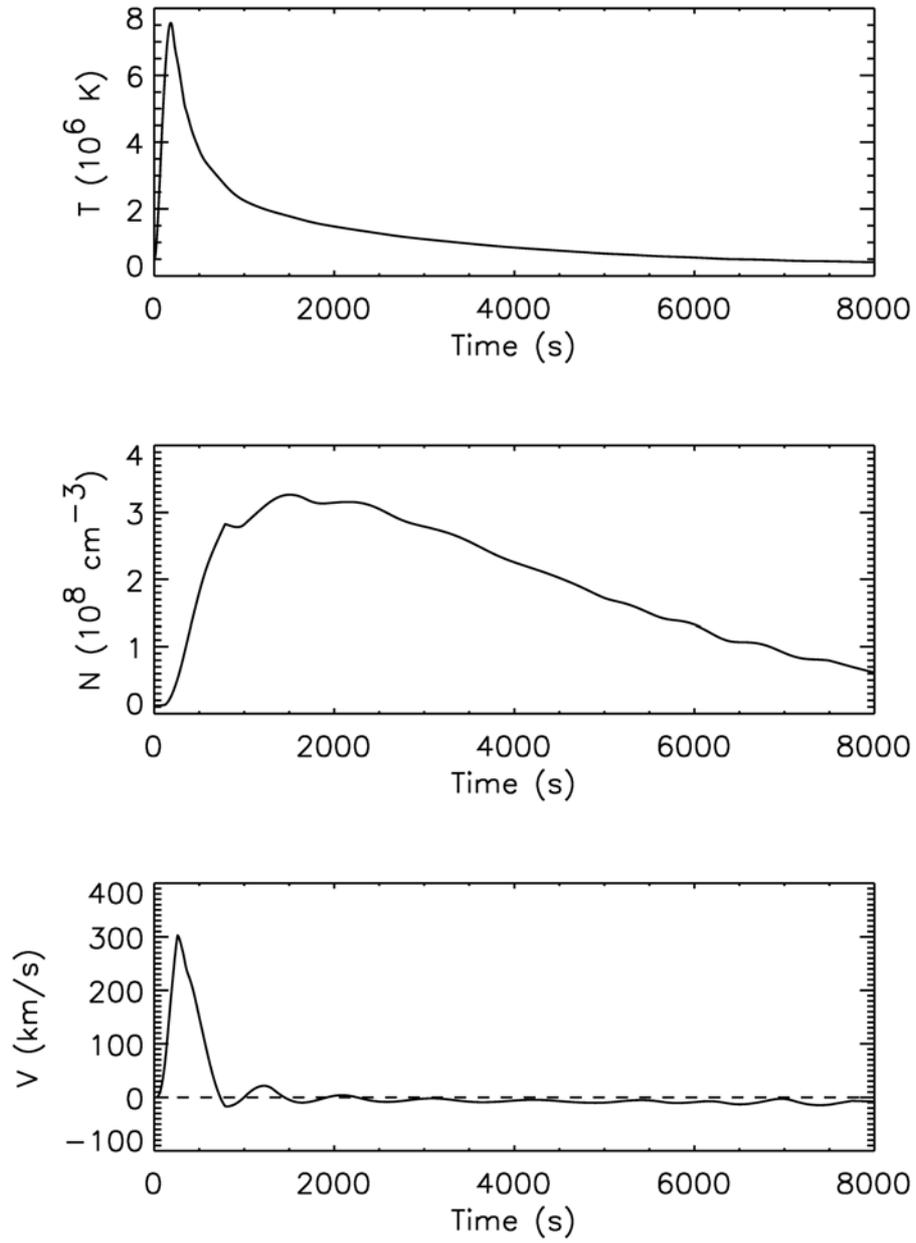

*Figure 2.* Spatially-averaged temperature, density, and velocity versus time for a loop strand that has been impulsively heated by a 500 s nanoflare. The averages are computed over the central 80% of the coronal section of the strand. Heat flux saturation is included.

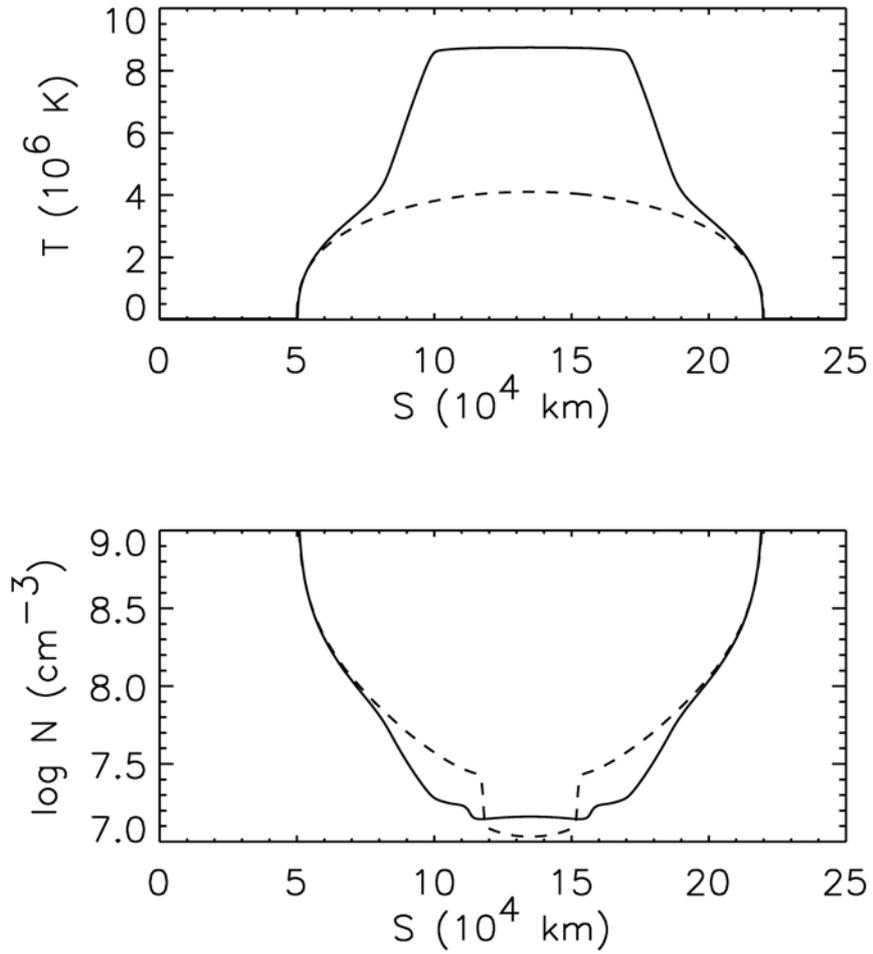

*Figure 3.* Temperature and density versus position along the strand at time *t* = 220 s, roughly halfway into the impulsive energy release. The solid and dashed curves are for simulations with and without heat flux saturation, respectively.

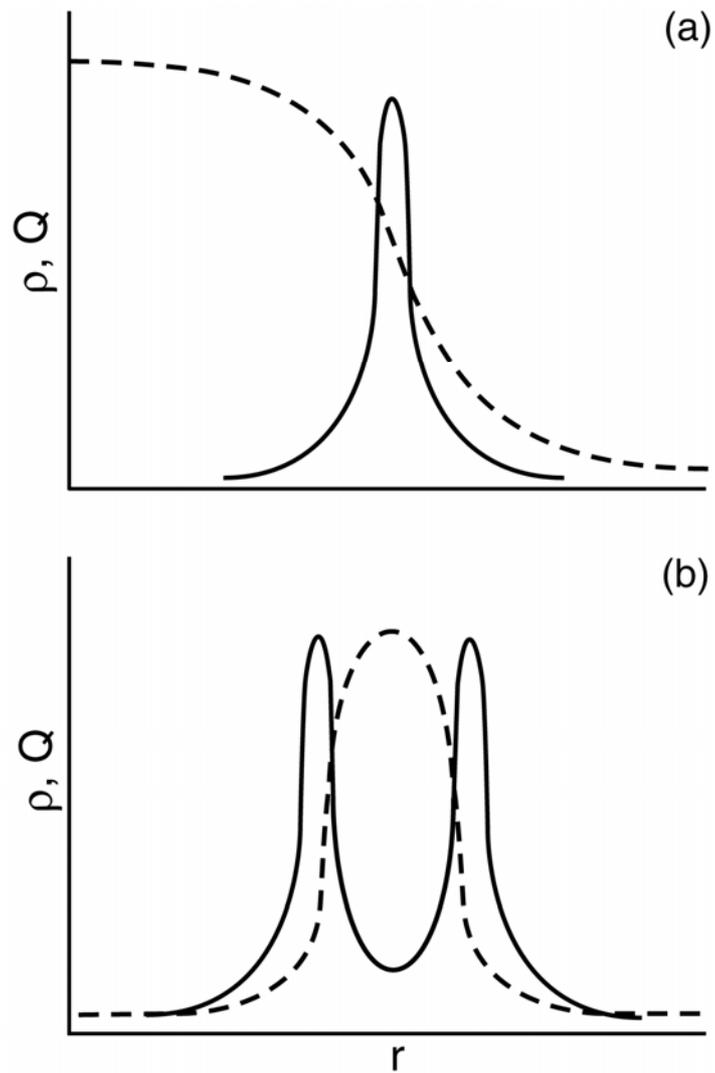

*Figure 4.* Schematic representation of resonant wave absorption. Density (dashed) and heating rate (solid) versus radial distance from the loop axis. Panel (a) corresponds to an initial time, and panel (b) corresponds to a subsequent time, as discussed in the text.

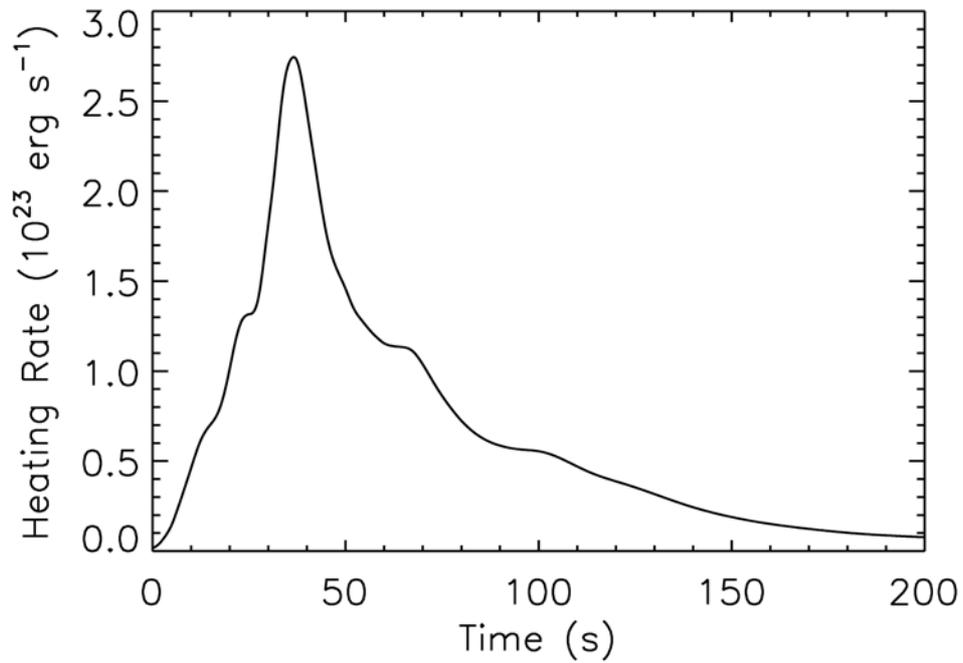

*Figure 5.* Combined resisitive and viscous heating rate versus time for a secondary instability simulation in which the magnetic field rotates by 90° across the current sheet (after Dahlburg, Klimchuk, and Antiochos, 2005).

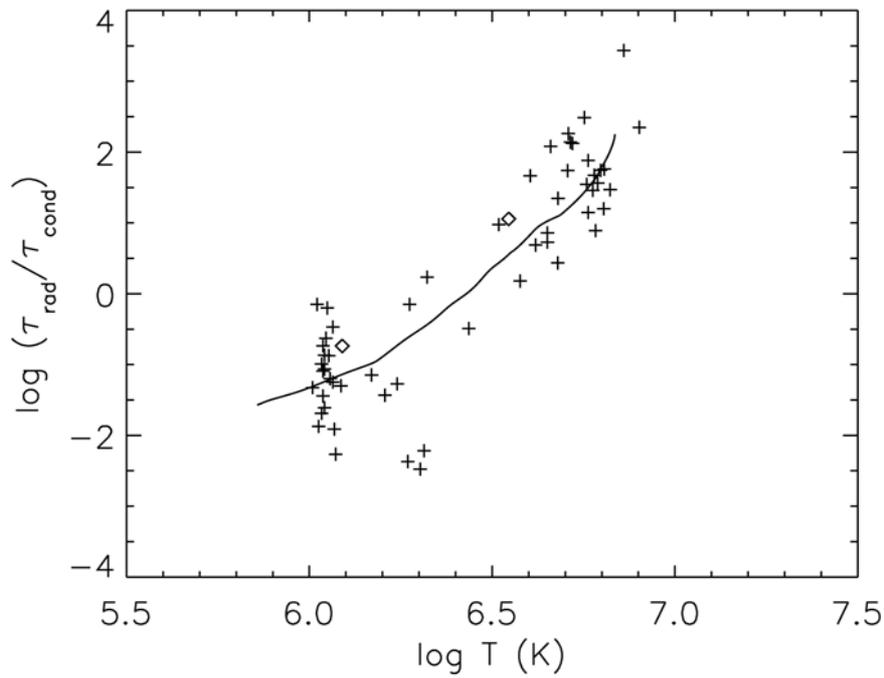

*Figure 6.* Ratio of radiative to conductive cooling times versus temperature for 56 different loops observed by either *Yohkoh* or *TRACE* (crosses). Solid curve shows the cooling track of a nanoflare simulation. Diamonds show synthetic observations based on the time-average of the simulation (*Yohkoh* observation at log*T* = 6.55; *TRACE* observation at log*T* = 6.09).

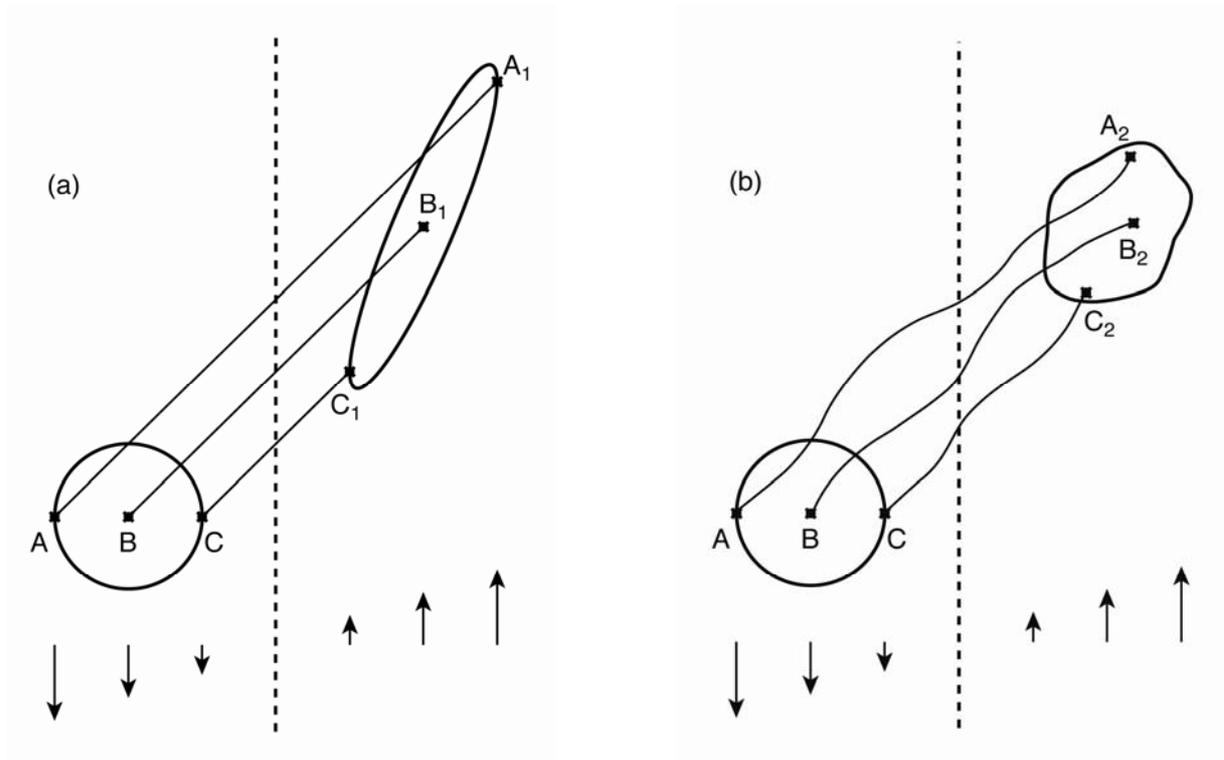

*Figure 7*. View from above of a coronal loop (a large magnetic flux tube) embedded in a translationally symmetric magnetic arcade. The arcade has been sheared by imposing anti-parallel boundary motions (arrows) on either side of the neutral line (dashed). The footpoint cross sections and three field lines, or elemental flux strands, are shown. The loop is defined by the circular cross section on the left, which is specified after the shear is applied. Panel (a) assumes that only the systematic shear flow is present, while panel (b) assumes an additional random component of flow and coronal reconnection.